\documentclass[fleqn,usenatbib]{mnras}

\usepackage{newtxtext,newtxmath}

\usepackage[T1]{fontenc}

\DeclareRobustCommand{\VAN}[3]{#2}
\let\VANthebibliography\thebibliography
\def\thebibliography{\DeclareRobustCommand{\VAN}[3]{##3}\VANthebibliography}

\usepackage{graphicx}	
\usepackage{amsmath}	

\usepackage[normalem]{ulem}
\usepackage{xcolor}
\usepackage{orcidlink}

\title[Disappearing supernova progenitors]{The disappearances of six supernova progenitors}

\author[S. D. Van Dyk et al.]{
Schuyler D.~Van Dyk,\orcidlink{0000-0001-9038-9950}$^{1}$\thanks{E-mail: vandyk@ipac.caltech.edu (SVD)}
Asia de Graw,$^{2}$
Raphael Baer-Way,$^{2}$
WeiKang Zheng,\orcidlink{0000-0002-2636-6508}$^{2}$
Alexei V.~Filippenko,\orcidlink{0000-0003-3460-0103}$^{2}$
\newauthor Ori D.~Fox,\orcidlink{0000-0003-2238-1572}$^{3}$ Nathan Smith,\orcidlink{0000-0001-5510-2424}$^{4}$ Thomas G.~Brink,\orcidlink{0000-0001-5955-2502}$^{2}$ Thomas de Jaeger,\orcidlink{0000-0001-6069-1139}$^{2,5}$ Patrick L.~Kelly\orcidlink{0000-0003-3142-997X}$^{6}$ and \newauthor Sergiy S.~Vasylyev\orcidlink{0000-0002-4951-8762}$^{2}$
\\
$^{1}$Caltech/IPAC, Mailcode 100-22, Pasadena, CA 91125, USA\\
$^{2}$Department of Astronomy, University of California, Berkeley, CA 94720-3411, USA\\
$^{3}$Space Telescope Science Institute, 3700 San Martin Drive, Baltimore, MD 21218, USA\\
$^{4}$Steward Observatory, University of Arizona, 933 North Cherry Avenue, Tucson, AZ 85721, USA\\
$^{5}$Institute for Astronomy, University of Hawai`i, 2680 Woodlawn Dr., Honolulu, HI 96822, USA\\
$^{6}$University of Minnesota, School of Physics and Astronomy, 116 Church St. SE, Minneapolis, MN 55455, USA
}

\date{Accepted 2022 November 29. Received 2022 November 28; in original form 2022 September 16}

\pubyear{2022}

\begin{document}
\label{firstpage}
\pagerange{\pageref{firstpage}--\pageref{lastpage}}
\maketitle

\begin{abstract}
As part of a larger completed {\sl Hubble Space Telescope\/} ({\sl HST}) Snapshot program, we observed the sites of six nearby core-collapse supernovae (SNe) at high spatial resolution: SN 2012A, SN 2013ej, SN 2016gkg, SN 2017eaw, SN 2018zd, and SN 2018aoq. These observations were all conducted at sufficiently late times in each SN's evolution to demonstrate that the massive-star progenitor candidate identified in each case in pre-explosion imaging data had indeed vanished and was therefore most likely the actual progenitor. However, we have determined for SN 2016gkg that the progenitor candidate was most likely a blend of two objects: the progenitor, which itself has likely vanished, and another closely-neighbouring star. We thus provide a revised estimate of that progenitor's properties: a binary system with a hydrogen-stripped primary star at explosion with effective temperature $\approx 6300$--7900\,K, bolometric luminosity $\approx 10^{4.65}$\,L$_{\odot}$, radius $\approx 118$--154\,$R_{\odot}$, and initial mass 9.5--11\,M$_{\odot}$. Utilising late-time additional archival {\sl HST\/} data nearly contemporaneous with our Snapshots, we also show that SN 2017eaw had a luminous ultraviolet excess, which is best explained as a result of ongoing interaction of the SN shock with pre-existing circumstellar matter. We offer the caveat, particularly in the case of SN 2013ej, that obscuration from SN dust may be compromising our conclusions. This sample adds to the growing list of confirmed or likely core-collapse SN progenitors.
\end{abstract}

\begin{keywords}
stars: massive -- stars: evolution -- supergiants -- binaries: general -- supernovae: individual -- SN 2012A, SN 2013ej, SN 2016gkg, SN 2017eaw, SN 2018zd, SN 2018aoq
\end{keywords}


\section{Introduction}\label{sec:intro}

The explosive endpoints of stellar evolution, supernovae (SNe) are among the most energetic events in the Universe. A consensus is that explosions of white dwarfs with donor companions in binary systems lead to thermonuclear Type Ia SNe. The terminal event in the lives of stars with zero-age main sequence (ZAMS) masses $M_{\rm ZAMS} \gtrsim 8$ M$_{\odot}$ is the collapse of the stellar core at the end of nuclear burning --- such core-collapse SNe (CCSNe) are observed as Type Ib, Ic, and II. (See \citealt{Filippenko1997} for a review of SN classification.) In the local Universe, CCSNe constitute the overwhelming majority, $\sim$76\%, of all SNe \citep{Li2011,Graur2017}. Among CCSNe, the Type II SNe (SNe~II) have classically been further divided into II-Plateau (II-P) and II-Linear (II-L; although the actual existence of a II-L distinction has been recently questioned, e.g., \citealt{Anderson2014,Valenti2016}); SNe II-P are $\sim$48\% of all CCSNe \citep{Smith2011}. A minority ($\sim$14\%) of all SNe~II appear to be dominated by interaction of the SN shock with a pre-existing circumstellar medium (CSM) and manifest themselves spectroscopically as II-narrow \citep[IIn; e.g.,][]{Schlegel1990,Smith2017}. The remainder of the CCSNe are the so-called ``stripped-envelope'' SNe (SESNe), in which the outer H-rich envelope of the progenitor star has been greatly or entirely removed, and for some SESNe even the He layer has been stripped away, prior to explosion. Included among the SESNe are the SNe~IIb ($\sim$10\% of all CCSNe; \citealt{Smith2011,Shivvers2017}), for which the progenitor retains a mass of hydrogen $M_H \lesssim 0.15$ M$_{\odot}$ at the time of its demise (\citealt{Yoon2017}; although more-extended SN~IIb progenitors may exceed this limit).

Mapping the various SN types to their progenitor systems is one of the primary goals of SN astrophysics. This knowledge can be acquired in various indirect ways, such as modeling SN light curves via progenitor population synthesis (e.g., \citealt{Eldridge2018}), probing the SN ejecta at late times (e.g., \citealt{Milisavljevic2012,Jerkstrand2012}), analysing the emission from the CSM ionized by early X-ray/ultraviolet (UV) radiation from the SN explosion (e.g., \citealt{Khazov2016}), and inferring ages and therefore turn-off masses from the SN's local stellar environment (e.g., \citealt{Maund2017,Williams2018,Corgan2022}). However, the most direct means of obtaining insight into the progenitor-SN connection is the precise identification of the star itself as seen at some time before the explosion  \citep{Smartt2009,Smartt2015,VanDyk2017b}. Whilst a few progenitors have been located using ground-based data, including SN 1978K \citep{Ryder1993}, SN 1987A (e.g., \citealt{Sonneborn1987}), SN 1993J \citep{Richmond1993,Aldering1994}, SN 2004et \citep{Li2005}, and SN 2008bk \citep{Mattila2008,VanDyk2012}, the majority have been identified in serendipitous pre-explosion imaging obtained with the {\sl Hubble Space Telescope\/} ({\sl HST}; \citealt{VanDyk2017a}, and references therein).

However, the identified star, or stellar system, remains merely a progenitor {\em candidate}, until its confirmation as the true progenitor through its disappearance. The key to this pursuit is waiting until the SN has faded sufficiently so one can successfully determine that the remaining light is less luminous than the original luminosity of the progenitor. This anticipated fading can be curtailed, though, if the SN is still interacting with a CSM, leading to excess luminosity above what would be expected from the gradual exponential decline rate associated with the reprocessing of $\gamma$-rays and positrons from radioactive decay. To date, the community has confirmed the progenitors of SN 1987A \citep{Gilmozzi1987,Walborn1987,Sonneborn1987}; SN 1993J and SN 2003gd \citep{Maund2009}; SN 2004A, SN 2005cs, and SN 2006my \citep{Maund2014}; SN 2005gl \citep{GalYam2009}; SN 2008ax \citep{Folatelli2015}; SN 2008bk \citep{Mattila2010,VanDyk2013a}; SN 2008cn \citep{Maund2015}; SN 2009ip \citep{Smith2022}; SN 2010bt \citep{EliasRosa2018}; SN 2011dh \citep{VanDyk2013b}; SN 2012aw \citep{VanDyk2015,Fraser2016}; iPTF13bvn \citep{Folatelli2016,Eldridge2016}; SN 2015bh \citep{Jencson2022}; and AT 2016jbu \citep{Brennan2022}. On the other hand, observing at sufficiently late times has also revealed that some candidates likely were {\em not\/} the progenitors at all, including SN 1999ev \citep{Maund2014}; SN 2006ov \citep{Crockett2011}; SN 2009kr and SN 2009md \citep{Maund2015}; and, SN 2010jl \citep{Fox2017}.

Of course, it is possible that, rather than the progenitor candidate outright vanishing, dust has formed and accumulated after the explosion and is obscuring the remaining emission from the SN. This would be particularly relevant if the progenitor candidate had experienced and survived a non-terminal explosion and was then obscured by the dust. In order to eliminate this possibility, observations of the SN would have to be undertaken in the infrared (IR). This has generally not been done for the previously-mentioned SNe. However, in a number of objects, such as SN 2003gd, the mass of dust estimated to be present at late times is quite low ($4 \times 10^{-5}$ M$_{\odot}$; \citealt{Meikle2007}) based on {\sl Spitzer Space Telescope\/} IR observations. For other SNe this may not necessarily be the case; we therefore offer this as a caveat on results of this kind. Also, unfortunately very few progenitor candidates have had pre-explosion IR counterparts isolated with which to compare directly. This situation may well change with the advent of the {\sl James Webb Space Telescope\/} ({\sl JWST}).

In this paper, as part of an {\sl HST\/} observing program that we had executed, we report on the apparent disappearances of six recent SNe (SNe 2012A, 2013ej, 2016gkg, 2017eaw, 2018zd, and 2018aoq), adding to the growing list of confirmed progenitors. This sample includes two normal SNe II-P, one low-luminosity SN II-P, one luminous possible SN II-P/II-L hybrid, one possible electron-capture event (or another SN~II-P), and one SN~IIb.

\section{Observations and Reductions}\label{sec:observations}

Observations were successfully conducted with {\sl HST\/} as a Cycle 28 Snapshot program, GO-16179 (PI A.~Filippenko). The sole instrument utilised during this program was the Wide Field Camera 3 (WFC3) with the UV/Visible channel (UVIS). A total of 37 visits were executed (out of 54 requested), in which data were acquired in two photometric bands within a single orbit per visit. A small line dither was employed between the two short exposures (frames) in each of the two bands, in an attempt to mitigate against cosmic-ray (CR) hits and imaging-array imperfections. A full description of the program and the complete results from it will be presented in a forthcoming paper. We note that no Exclusive Access period was imposed on the program's data, which were public as soon as they were posted in the Mikulski Archive for Space Telescopes (MAST). Here we highlight the results from the six visits that included the SNe mentioned above. Several of these SNe were expressly targeted in order to determine whether the progenitors had indeed vanished.

The two frames per band per visit were mosaicked using the routine {\tt Astrodrizzle} \citep{STScI2012}. An important outcome from running this routine is that CR hits are masked in the data quality (DQ) array of each frame. All of the frames per visit were then processed with {\tt Dolphot} \citep{Dolphin2016}, in order to extract photometry via point-spread-function (PSF) fitting of all of the stellar (or stellar-like) objects detected above a nominal 3$\sigma$ threshold in the mosaic of one band that is used as a reference image. We adopted the {\tt Dolphot} input parameters FitSky=3, RAper=8, and InterpPSFlib=1 (since the UVIS frames have already been corrected in the standard pipeline for charge-transfer efficiency, we set WFC3useCTE=0). The photometric results from {\tt Dolphot} are returned in Vega magnitudes. We then employed prior images, which in most cases were previous {\sl HST\/} images of the SN or of its progenitor, to precisely isolate the late-time counterpart of the SN in a Snapshot visit. In some cases below, {\tt Dolphot} did not detect a counterpart at the SN position; even a visual inspection in these cases readily revealed that the SN was no longer detectable to the exposure depth of the data.

Calculating upper limits in these instances is not straightforward. After some consideration we have chosen to set the upper limits based on the formal signal-to-noise ratio (S/N) provided by {\tt Dolphot}. However, we point out that \citet{Williams2014} found that {\tt Dolphot} tends to underestimate photometric uncertainties, particularly in crowded environments; thus, setting accurate limits, based on S/N, at the lowest flux levels is hampered by this underestimation. Similarly, we have since found that injecting artificial stars with {\tt Dolphot} at a SN site at flux levels near the image noise floor (e.g., \citealt{VanDyk2016}) tends to lead to recovered fluxes that are systematically too high (possibly as a result of the PSF fitting including too much background around the injected star). Fortunately, the locations for which we need to impose upper limits are relatively uncrowded environments, and the photometric uncertainties are probably only underestimated by factors of a few \citep{Williams2014}. To be conservative, therefore, we hereafter set upper limits at S/N $=5$ based on photometry around the SN site, with the realisation that these limits could well be at S/N $\approx 1.7$--2.5. The reader should take all of these limitations and caveats into account when we quote detection upper limits in this paper.

\section{Discussion of Individual Supernovae}\label{sec:individual}

\subsection{SN 2012A}

\citet{Tomasella2013} demonstrated through detailed optical/IR photometric and spectroscopic monitoring of SN 2012A in NGC 3239 that it was a normal SN II-P. We now have overwhelming evidence that SNe II-P, as theoretically expected, arise from massive stars in the red supergiant (RSG) phase, with $M_{\rm ZAMS}$ in the range of $\sim$8--17 M$_{\odot}$ \citep{Smartt2009,Smartt2015}, although arguments have been made that the range may extend up to $\sim$19 M$_{\odot}$ or higher \citep{Davies2018}. In the case of SN 2012A, \citet{Prieto2012} first identified a point source at the SN position in high-resolution pre-explosion Gemini-North Near-InfraRed Imager and Spectrometer (NIRI) $K'$ images. The detection was only in this one single band. Their preliminary estimate of the apparent brightness of the progenitor candidate was $K = 20.1 \pm 0.2$\,mag. Adjusting this measurement to an assumed distance to the host galaxy resulted in an absolute brightness $M_K \approx -9.4$\,mag, consistent with expectations for RSGs.

\citet{Tomasella2013} reanalysed the Gemini data and, photometrically calibrating with 2MASS $K_s$ (similarly to \citealt{Prieto2012}), found that the candidate had $K^{\prime}$= $20.29 \pm 0.13$\,mag. Given the distance to the host galaxy and the $K$-band extinction they assumed, as well as an estimated bolometric correction for $K$, they concluded that the candidate had a bolometric magnitude $-7.08 \pm 0.36$ and thus a bolometric luminosity $\log (L/{\rm L}_{\odot}) = 4.73 \pm 0.14$. From a comparison to theoretical single-star evolutionary tracks, \citet{Tomasella2013} concluded that the RSG star had $M_{\rm ZAMS} = 10.5^{+4.5}_{-2}$ M$_{\odot}$. As those authors noted, their treatment of the progenitor candidate had not included possible circumstellar dust in the RSG envelope, although they reasoned that any circumstellar extinction in the optical would be less by about a factor of $\sim$10 at $K$, so the impacts of dust should be minimal on their luminosity estimate for the star.

\begin{figure*}
	\includegraphics[width=\columnwidth]{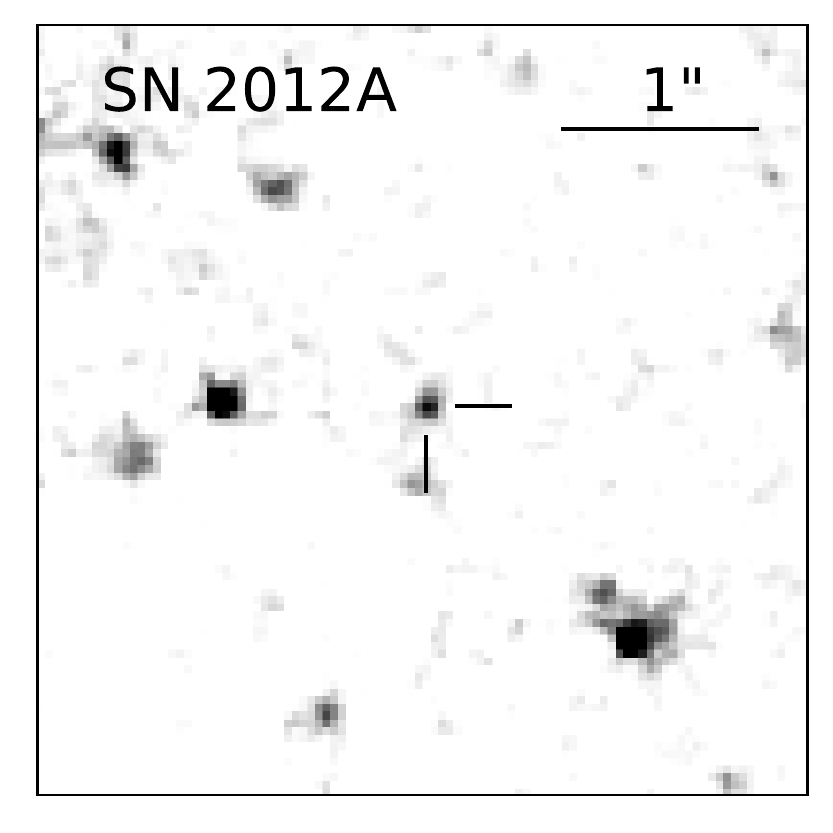}
	\includegraphics[width=\columnwidth]{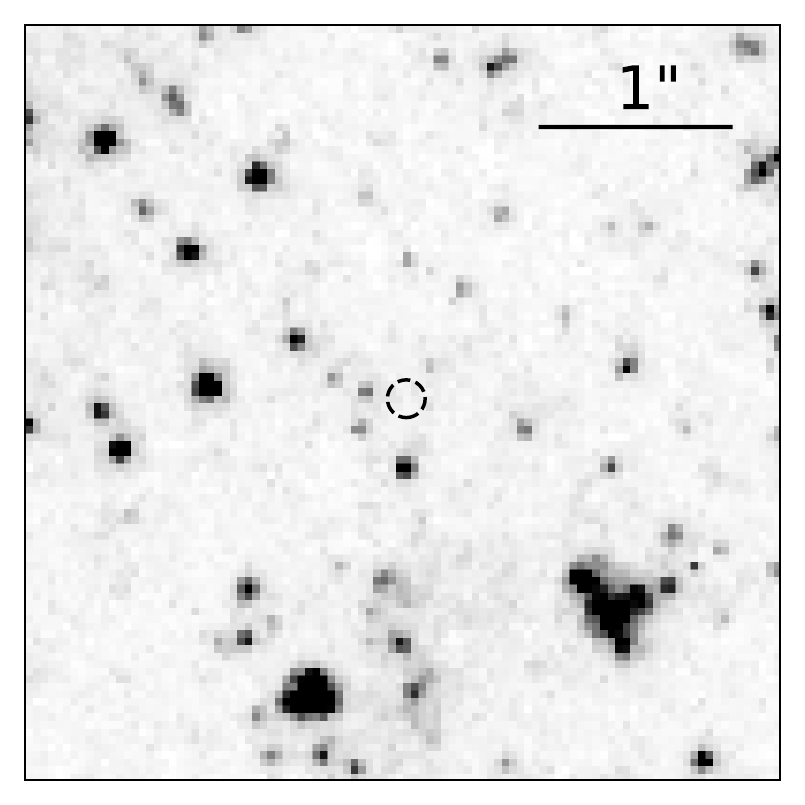}
    \caption{{\it Left}: A portion of the pre-explosion Gemini-N NIRI $K'$-band mosaic from 2006 May 13, with the progenitor candidate for SN 2012A indicated by tick marks (see also~\citealt{Tomasella2013}, their Figure 16). {\it Right}: A portion of the WFC3/UVIS F814W mosaic from 2021 February 16, with the corresponding position of the progenitor candidate encircled. No source is detected at the SN position to $> 25.8$\,mag in that band. North is up, and east is to the left.}
    \label{fig:sn2012A}
\end{figure*}

The {\sl HST\/} observations of SN 2012A from our program in F606W and F814W occurred on 2021 February 16 (UT dates are used throughout this paper), 3329\,d ($\sim 9.1$\,yr) after explosion \citep{Tomasella2013}. The total exposure times were 710 s (F606W) and 780 s (F814W). (Total exposure times will be listed hereafter in parentheses.) We focus on the F814W band, since it is the closer of the two Snapshot bands to $K'$ in wavelength. A comparison between $K'$ and F814W is shown in Figure~\ref{fig:sn2012A}. For the purposes of the figure, we reconstructed the $K'$ mosaic from the original NIRI data (in a similar fashion to that of \citealt{Tomasella2013}). To locate the SN position in the F814W image, we used 12 stars in common to astrometrically register the two images, with a $1\sigma$ root-mean-square (RMS) uncertainty of 0.19 UVIS pixel (7.5 mas). We estimated an upper limit at that position of $> 25.8$\,mag in F814W (for completeness we note an upper limit in F606W of $> 26.9$\,mag). For context we further extrapolated the observed exponential decline in the $I$-band light curve from \citet{Tomasella2013} to the SN 2012A Snapshot observation date and would expect the SN to have been at $I \approx 55$\,mag (!). We can therefore fairly safely rule out any residual SN emission. We note that we have considered the upper limit at F814W, but the actual upper limit at $K'$ would depend on the expected ${\rm F814W}-K'$ colour for an RSG. Assuming an RSG with effective temperature $T_{\rm eff}=3600$ K, and including the assumed extinction to SN 2012A, the colour of the star would be $\sim 2.48$\,mag, such that the presumed limit on the progenitor candidate is then $K' \gtrsim 23.3$\,mag. If dust from the SN were obscuring the candidate, this would require an extinction $A_K \gtrsim 3.0$\,mag ($A_V \gtrsim 27.3$\,mag [!]) from such dust. We consider it far more likely that the putative RSG progenitor candidate has vanished.

\subsection{SN 2013ej}

SN 2013ej in NGC 628 (M74) drew immediate attention, owing to its relative proximity (9.8\,Mpc) and its occurrence in a host of multiple SNe. Several studies presented results of intensive UV/optical/IR photometric and spectroscopic monitoring (e.g., \citealt{Valenti2014,Bose2015,Huang2015,Dhungana2016,Yuan2016}). The unusually long rise, comparatively shorter and steeper plateau phase, high luminosity, and spectral indications of CSM interaction were highly suggestive of an unusual SN~II, with characteristics pointing toward the classical II-L designation. At the very least, it appears it could be considered an atypical SN II-P or an SN II-P/II-L hybrid, as some in the community have already been tagging it. \citet{Chakraborti2016} observed and modeled X-ray emission from SN 2013ej and found evidence for steady mass loss during the final 400\,yr of the progenitor star. Polarimetry, both photometric \citep{Kumar2016} and spectroscopic \citep{Mauerhan2017,Nagao2021}, have pointed to a complex circumstellar environment around the SN.

\citet{Fraser2014} identified a progenitor candidate in pre-explosion {\sl HST\/} images obtained with the Advanced Camera for Surveys (ACS)/Wide-Field channel (WFC) in bands F435W, F555W, and F814W on 2003 November 20, 2003 December 29 (F555W only), and 2005 June 16. What those authors found was that SN 2013ej was most coincident with the source at F814W, at $22.66 \pm 0.03$\,mag (\citealt{Mauerhan2017} measured a slightly brighter $22.60 \pm 0.01$ and $22.57 \pm 0.02$\,mag from 2003 and 2005, respectively), but that the photocentres of the source at both F435W and F555W were significantly offset (by $\sim$40--47\,mas, an $8\sigma$ displacement; at the M74 distance, this is $\sim 2$\,pc) relative to F814W, implying that the source detected at F435W and F555W is unrelated to the one at F814W, which they ascribed to the SN progenitor. (Lead author S.~Van Dyk had conducted a similar unpublished analysis and found essentially the same results.) The unrelated source was at $25.14 \pm 0.07$ and $24.98 \pm 0.05$\,mag in F435W and F555W, respectively. (Wide Field and Planetary Camera 2 [WFPC2] F336W data also exist, with the detected source at $23.31 \pm 0.14$\,mag.) Assuming that the candidate was an RSG, \citet{Fraser2014}, based on F814W only, concluded that its luminosity was in the range $\log (L/{\rm L}_{\odot}) = 4.46$--4.85; comparing to the endpoints of stellar evolution models, $M_{\rm ZAMS}=8$--15.5\,M$_{\odot}$, or at the least, $<16$ M$_{\odot}$. We note that, based on the photometric and spectroscopic observations, the progenitor mass was estimated to be $M_{\rm ZAMS} \approx 14$ M$_{\odot}$, 12--13\,M$_{\odot}$, $\sim$15 M$_{\odot}$, and 12--15\,M$_{\odot}$ by \citet{Bose2015}, \cite{Huang2015}, \citet{Dhungana2016}, and \citet{Yuan2016}, respectively.

\begin{figure*}
	\includegraphics[width=\columnwidth]{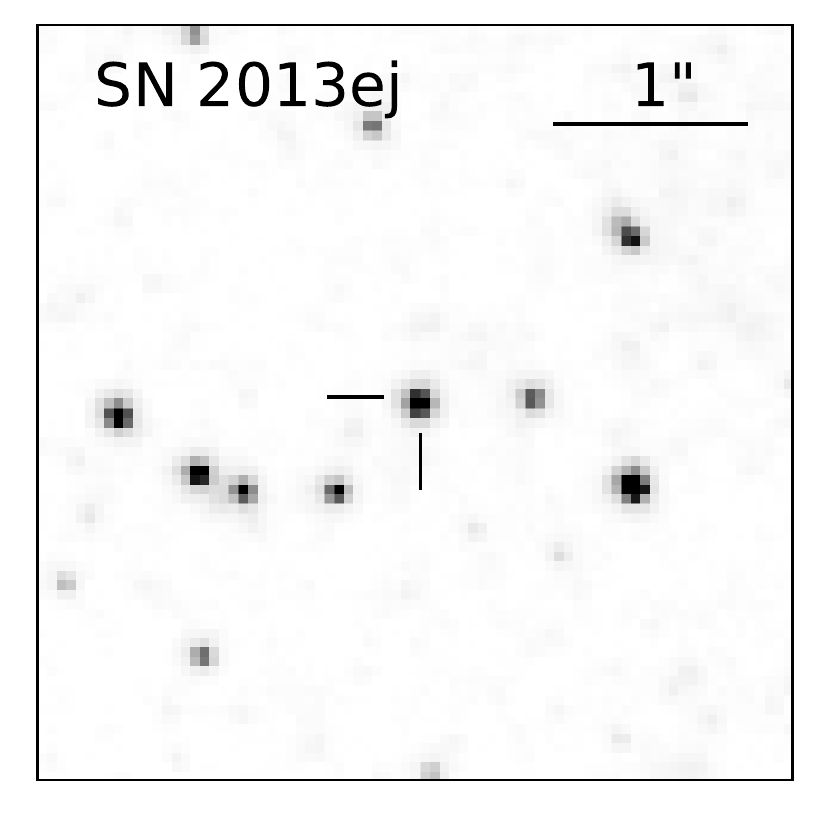}
	\includegraphics[width=\columnwidth]{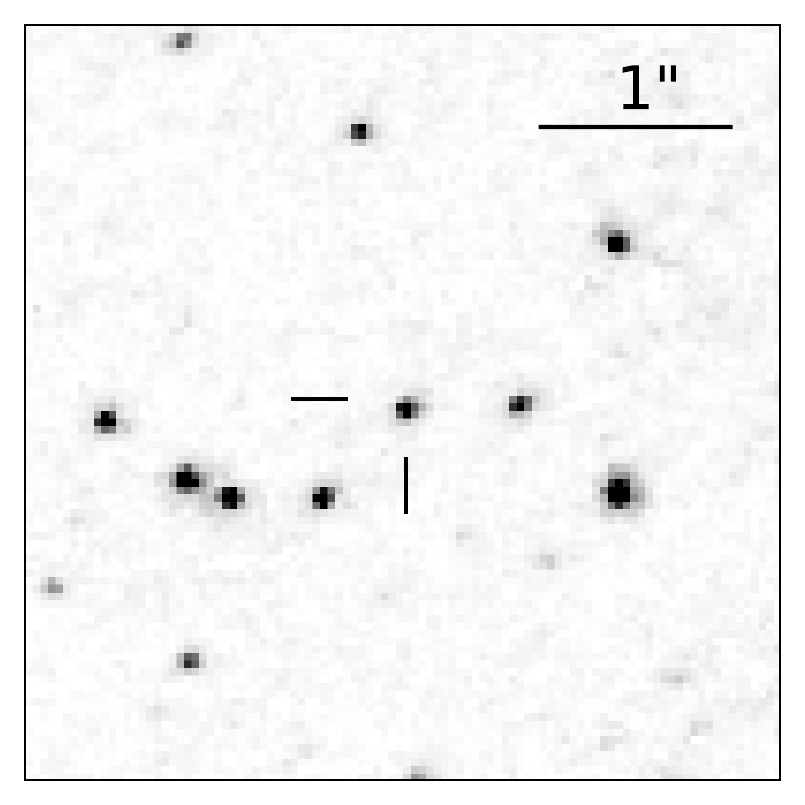}
    \caption{{\it Left}: A portion of the pre-explosion ACS/WFC F814W mosaic from 2003 November 20, with the progenitor candidate for SN 2013ej indicated by tick marks (see also \citealt{Fraser2014}, their Figure 1). {\it Right}: A portion of the WFC3/UVIS F814W mosaic from 2021 August 19, with the corresponding position of the progenitor candidate indicated by tick marks. A source is still clearly visible in the 2021 Snapshots in both F555W and F814W. North is up, and east is to the left.}
    \label{fig:sn2013ej}
\end{figure*}

The Snapshot observations of SN 2013ej were obtained on 2021 August 19 at F555W (710 s) and F814W (780 s), 2949.5\,d ($\sim$8.1 yr) after explosion. (Note that we had also obtained F438W and F625W snapshots, as part of the same {\sl HST\/} program, in 2021 February of the site of AT 2019krl, which is also in M74; however, the SN 2013ej site unfortunately fell within the chip gap for both of those bands.) One can see in Figure~\ref{fig:sn2013ej} that a source at the SN position is still clearly detected in the observations. We measured brightnesses of $24.26 \pm 0.03$ and $23.39 \pm 0.03$\,mag in the two respective bands. Comparing with the published values \citep{Fraser2014}, the light at the position of the progenitor candidate is $\sim$0.75 mag fainter in F814W than in 2003 or 2005. \citet{Mauerhan2017} reported that, based on WFC3/UVIS observations from 2015 December (GO-14116; PI S.~Van Dyk) and 2016 October (GO-14668; PI A.~Filippenko), the candidate was $\sim$0.46 mag fainter; however, we consider that the decline to the 2021 value is a more convincing decrease. We therefore believe we can now safely and strongly say that the candidate has vanished and was therefore the progenitor. 

Nevertheless, the light in F555W is still brighter, by $\sim$0.61 mag, than what was measured pre-explosion, implying that the SN itself is still contributing significantly, likely as a result of ongoing CSM interaction (e.g., the H$\alpha$ and [O~{\sc i}] emission spectroscopically detected by \citealt{Mauerhan2017} in 2016 are within the F555W bandpass). Furthermore, the SN image in F555W is offset by 0.38 UVIS pixel relative to F814W (compared to an astrometric 1$\sigma$ uncertainty of 0.15\,pixel), and the {\tt Dolphot} output parameter ``roundness'' is 0.353 at F555W, whereas it is 0.222 at F814W. So, we surmise that the light at F555W includes not only the old SN, but also light from the previously-mentioned contaminating neighbouring source or sources. Therefore, although the SN has now faded substantially in F814W, it is not yet faint enough in both bands that we can obtain a more detailed and less confused view of the immediate SN environment (within $\sim$1--2 pixels).

Before we can be certain about the SN 2013ej progenitor, though, we must note that this case is not nearly as clear-cut as for SN 2012A, above, owing to the likely presence of dust formed in a cold dense shell behind the reverse shock in the SN-CSM interaction region \citep{Mauerhan2017}. \citet{Mauerhan2017} observed a very late-time rebrightening in the mid-IR, based on {\sl Spitzer\/} data, between $\sim$700 and 1000 d after explosion (see also \citealt{Szalai2021}). Furthermore, the SN was clearly detected in early observations in 2022 with the {\sl JWST\/} NIRCam \citep{Pierel2022}. So, we cannot at this point neglect the effects of dust on the light that we measure in our Snapshot data. We therefore strongly encourage further monitoring of SN 2013ej with {\sl HST}, as well as with {\sl JWST}, for as long as possible.

\subsection{SN 2016gkg}

\bibpunct[;]{(}{)}{;}{a}{}{;}

SN 2016gkg in NGC 613 was discovered by an Argentinian amateur astronomer literally within a few hours after explosion, and was soon shown to be an SN IIb. \citet{Arcavi2017} modeled the very early shock cooling after the initial luminosity peak and determined that the progenitor had a radius of $\sim$40--150\,$R_{\odot}$ with $\sim 2$--$40 \times 10^{-2}$ M$_{\odot}$ in its extended envelope. \citet{Piro2017} also modeled the first cooling peak and found $\sim$180--260\,$R_{\odot}$ and $\sim 0.02$ M$_{\odot}$. The double-peaked light curve, characteristic of some SNe~IIb, became evident with continued early monitoring. \citet{Tartaglia2017} reported on the isolation of a progenitor candidate in {\sl HST\/}/WFPC2 F450W, F606W, and F814W images. Based on analysis of their photometry of what they considered ``Source A'' (the candidate), they inferred that the progenitor was of about mid-F spectral type and $M_{\rm ZAMS} = 15$--20\,M$_{\odot}$. \citet{Kilpatrick2017} had also identified the progenitor candidate in the same {\sl HST\/} data and characterised it as having a temperature of 9500~K and luminosity $\log (L/{\rm L}_{\odot}) = 5.15$, consistent with an A0~Ia class; their estimate of the radius was $\sim 257\,R_{\odot}$. Through Bayesian inference, \citet{Sravan2018} concluded the probability that the progenitor was a binary system was as high as 44\%, with a main-sequence or red-giant companion. At late times (300--800\,d), \citet{Kuncarayakti2020} found evidence from nebular spectroscopy for an asymmetric explosion with two-velocity-component ejecta, whilst radio emission detected up to 1429\,d provided evidence of SN shock-CSM interaction.

\citet{Bersten2018} also fully characterised both the SN properties and the progenitor. They found from their hydrodynamic modeling that an envelope radius of $\sim 320\,R_{\odot}$ with $\sim$ 0.01\,M$_{\odot}$, and ejecta mass $\sim$3.4 M$_{\odot}$, best reproduced the observed SN properties. Furthermore, numerical simulations of the properties of the progenitor --- $T_{\rm eff} \approx 7250$\,K and $\log (L/{\rm L}_{\odot}) \approx 5.10$ --- inferred from analysis of the pre-explosion photometry after unambiguously identifying the candidate, implied that the progenitor was a binary system consisting of components $M_{\rm ZAMS} = 19.5$ M$_{\odot}$ (primary) and 13.5\,M$_{\odot}$ (companion) with an initial orbital period of 70\,d. Those were the initial parameters of the binary; in the same model \citet{Bersten2018} proposed the stripped progenitor's mass at the time of explosion was reduced to 4.6 M$_{\odot}$, whereas the companion had gained mass through mass transfer and was likely a main sequence star with a mass of 17--20.5 M$_{\odot}$ (depending on the accretion efficiency) at the time of the progenitor's explosion, and the final orbital period had increased substantially to 631 d.

\begin{figure*}
	\includegraphics[width=\columnwidth]{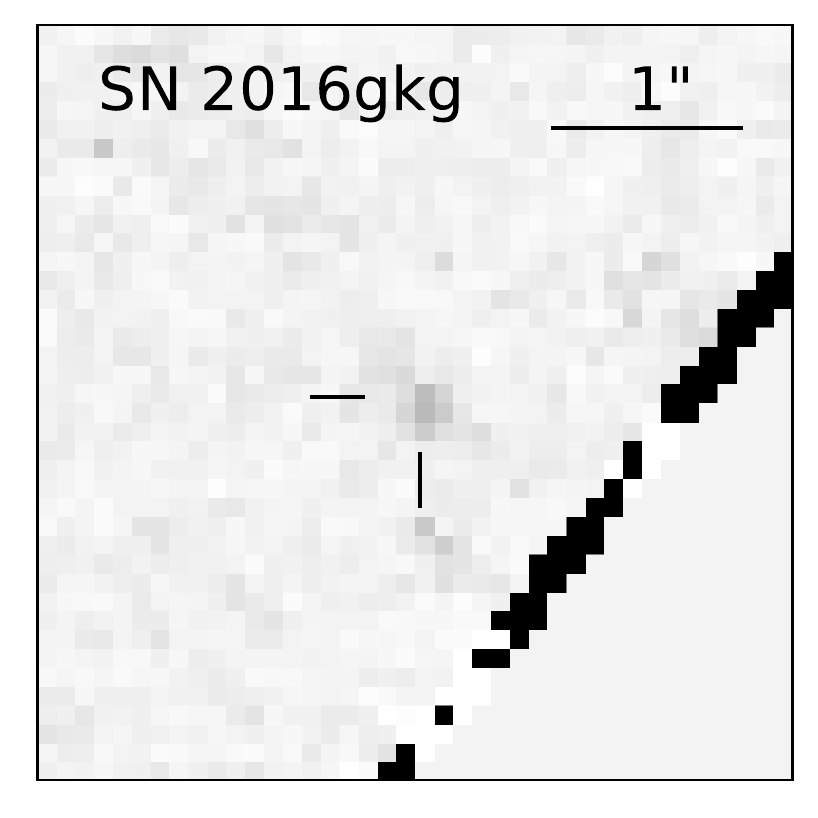}
	\includegraphics[width=\columnwidth]{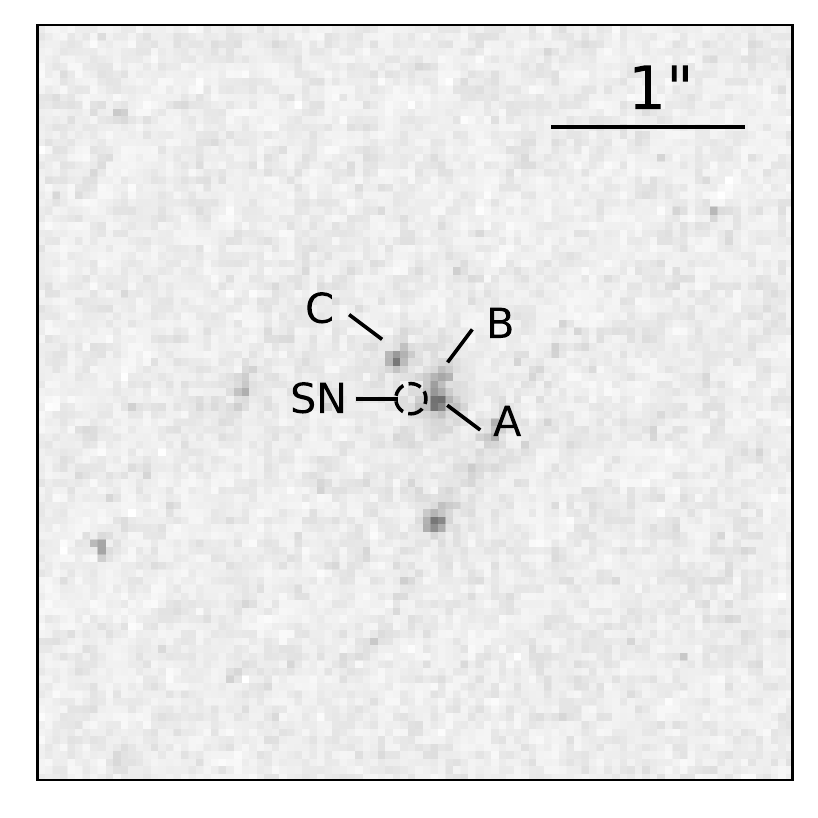}
    \caption{{\it Left}: A portion of the pre-explosion WFPC2 F450W mosaic from 2001 August 21, with the progenitor candidate for SN 2016gkg indicated by tick marks (see also, e.g., \citealt{Bersten2018}, their Extended Data Figure 6). The progenitor site was very near the edge of the mosaic. {\it Right}: A portion of the WFC3/UVIS F438W mosaic from 2021 August 19 (nearly 20\,yr later!), with the corresponding position of the progenitor candidate indicated by tick marks. We also identify three other resolved sources in the immediate environment, Stars ``A,'' ``B,'' and ``C.'' North is up, and east is to the left.}
    \label{fig:sn2016gkg_F438W}
\end{figure*}

\begin{figure*}
	\includegraphics[width=\columnwidth]{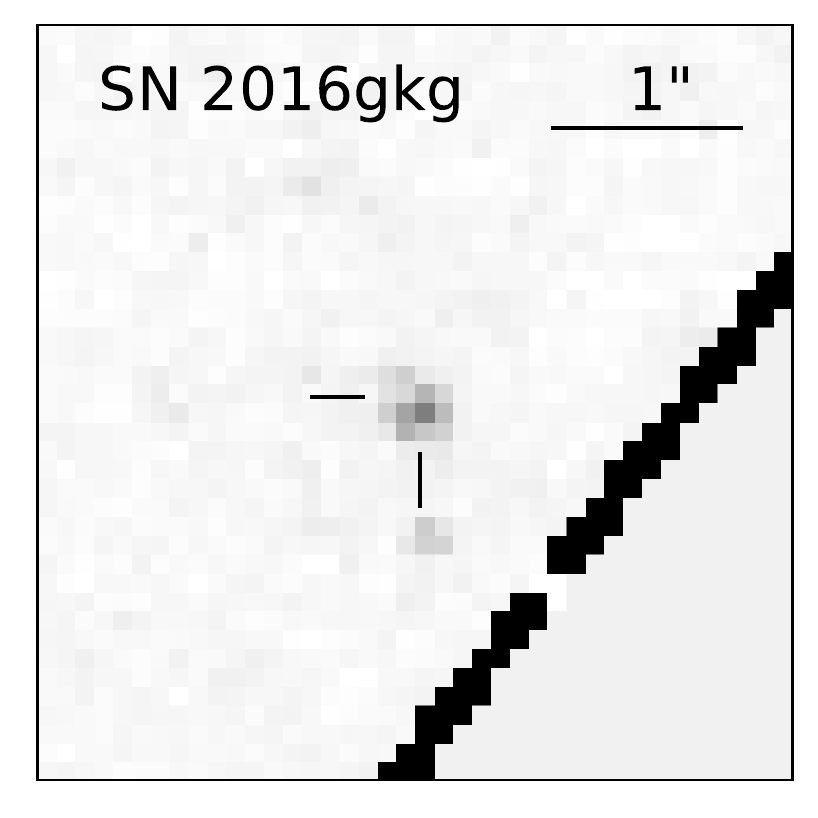}
	\includegraphics[width=\columnwidth]{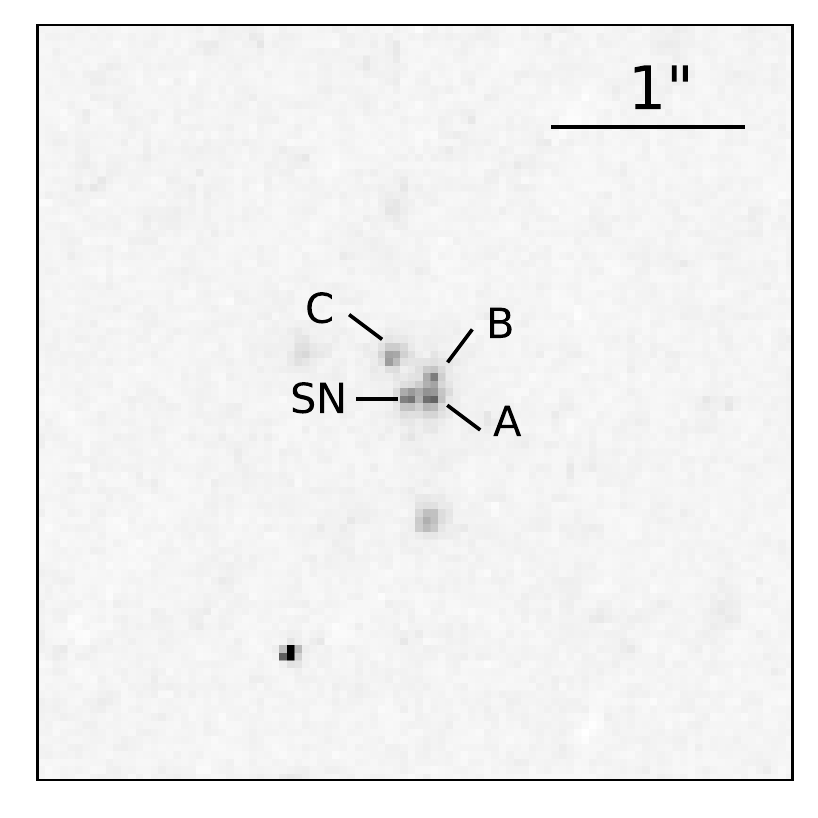}
    \caption{Same as Figure~\ref{fig:sn2016gkg_F438W}, but with {\it left}, WFPC2 F606W from 2001 August 21 and, {\it right}, WFC3/UVIS F606W mosaic from 2021 August 19.}
    \label{fig:sn2016gkg_F606W}
\end{figure*}

Our Snapshot observations in F438W (710 s) and F606W (780 s) occurred on 2021 August 19, 1794.7\,d (4.9\,yr) after explosion. We show in Figures~\ref{fig:sn2016gkg_F438W} and \ref{fig:sn2016gkg_F606W} the results compared to the pre-SN data. It is clear that the SN environment, as \citet{Tartaglia2017} had first found, is quite complicated. This is by far the most complex environment of any of the SNe that we consider in this paper. We were able to confidently identify which source was most likely associated with the SN, much as \citet{Bersten2018} had done, by astrometrically registering our Snapshots to the early-time 2016 WFC3/UVIS imaging of the SN in F555W (GO-14115; PI S.~Van Dyk) with a $1\sigma$ RMS uncertainty of 0.76 UVIS pixel ($0{\farcs}03$). We note that \citet{Kilpatrick2022} also analysed our Snapshot images and identified the same SN counterpart, although they arrived at slightly different values for the source's brightness, with somewhat larger uncertainties, i.e., $26.61 \pm 0.27$ and $25.10 \pm 0.07$\,mag in F438W and F606W, respectively; we measured the corresponding brightness of that source to be $>26.8$ and $24.88 \pm 0.04$\,mag. We therefore find the SN to be somewhat brighter in F606W than did \citet{Kilpatrick2022} and not detected at all in F438W. The sustained late-time brightness in F606W likely arises, as \citet{Kuncarayakti2020} and \citet{Kilpatrick2022} have shown, from strong [O~{\sc i}] $\lambda\lambda$6300, 6364 emission, and weaker H$\alpha$+[N~{\sc ii}] emission, within that band.

In the SN environment, as indicated in Figures~\ref{fig:sn2016gkg_F438W} and \ref{fig:sn2016gkg_F606W}, we identify three other objects, in addition to the SN itself; these are labelled as Stars A, B, and C. The first two stars, along with the SN, are confined to a tight cluster-like gathering over $\sim 0{\farcs}4$, whereas Star C is clearly separated from that cluster, and from the SN, by $\sim 0{\farcs}2$ (in our assessment, this object is likely ``Star B'' of \citealt{Tartaglia2017}). We measure brightnesses for the three stars in F438W and F606W (respectively) of $24.96 \pm 0.06$ and $24.85 \pm 0.04$ (A); $25.85 \pm 0.12$ and $25.14 \pm 0.05$ (B); and, $25.40 \pm 0.08$ and $25.04 \pm 0.05$\,mag (C).

We reprocessed the pre-explosion WFPC2 data for the host galaxy yet again with {\tt astrodrizzle} and {\tt Dolphot}, obtaining $24.00 \pm 0.15$ and $23.84 \pm 0.06$\,mag in F450W and F606W, respectively, for the progenitor candidate. These results agree quite well at F450W with those of \citet{Bersten2018}, although here we find the progenitor candidate to be somewhat brighter in F606W. (The main differences in this processing with that of \citealt{Bersten2018} is that here we set WFPC2UseCTE=1 and have also flagged CR hits prior to {\tt Dolphot}, which had not been done in the prior study.) We also very carefully inspected the residual images created after the PSF fitting photometry. By comparing to our Snapshot data, the progenitor candidate in WFPC2, as extracted in the PSF fitting, appears to be a blend of the profiles of the progenitor and Star A --- {\tt Dolphot} detects Star C as a separate object and does not detect Star B at all. Although we cannot be certain that the blend did not also include at least some of the light from Star B or from other, fainter objects that were not detectable, if we subtract the brightness of Star A from the progenitor candidate, we infer that the brightness of the progenitor alone would have been $24.58 \pm 0.22$ and $24.38 \pm 0.22$\,mag in F450W and F814W, respectively\footnote{With the compounding effects of a $5\sigma$ detection limit of $24.8$\,mag in F450W, and a $\sim 1.3$\,mag difference in brightness and separation of only $\sim 0{\farcs}015$ from the progenitor, it is not surprising that Star B was not detected in the WFPC2 data.}. We note that this is in overall agreement, to within the uncertainties, with the brightness found by \citet{Kilpatrick2022} based on forced photometry at the progenitor site. 

If this inference is correct, then we can certainly say that, based on our measurements from the Snapshot data, the progenitor candidate for SN 2016gkg has vanished. \citet{Kilpatrick2022} came to a similar conclusion.

\begin{figure*}
	\includegraphics[width=\columnwidth]{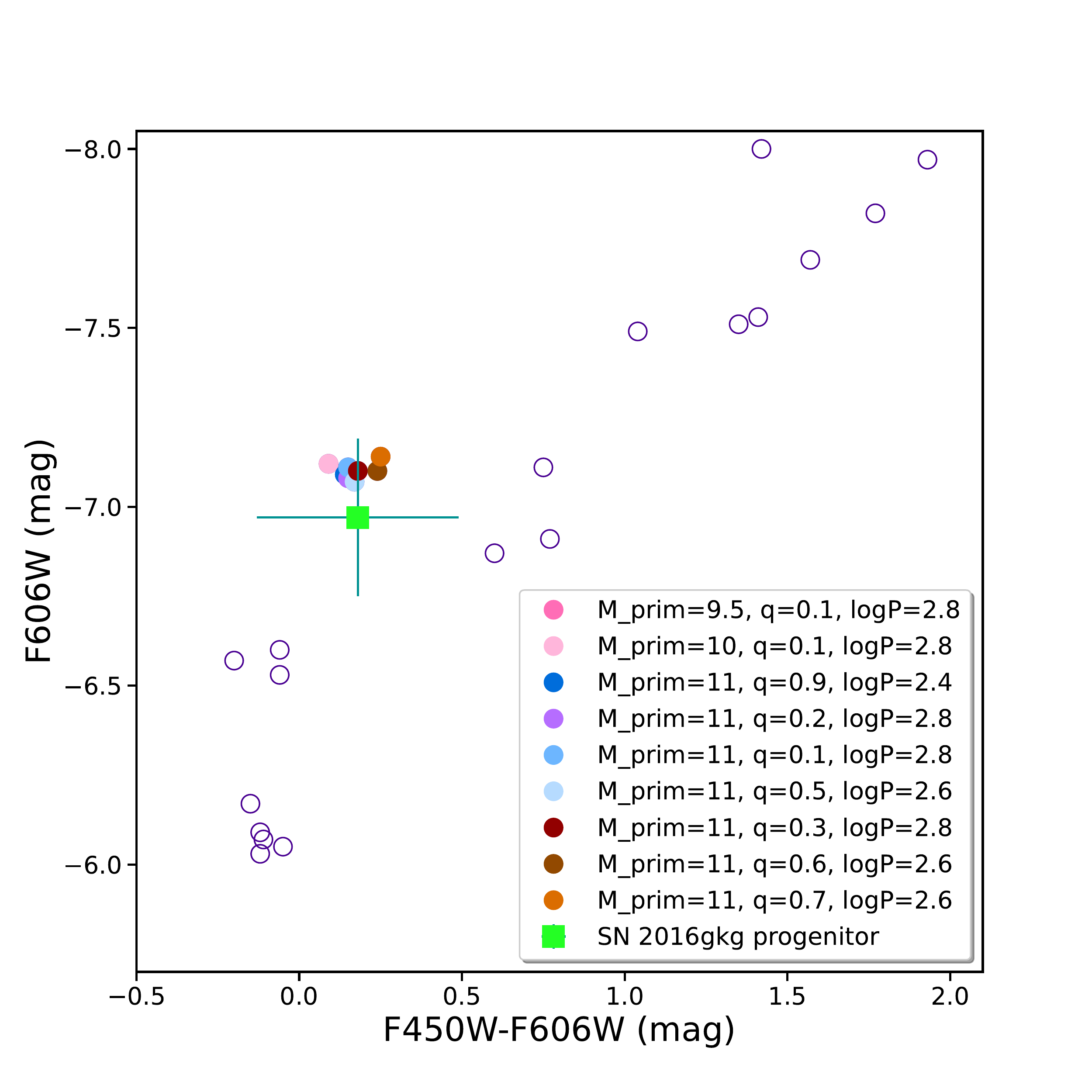}
	\includegraphics[width=\columnwidth]{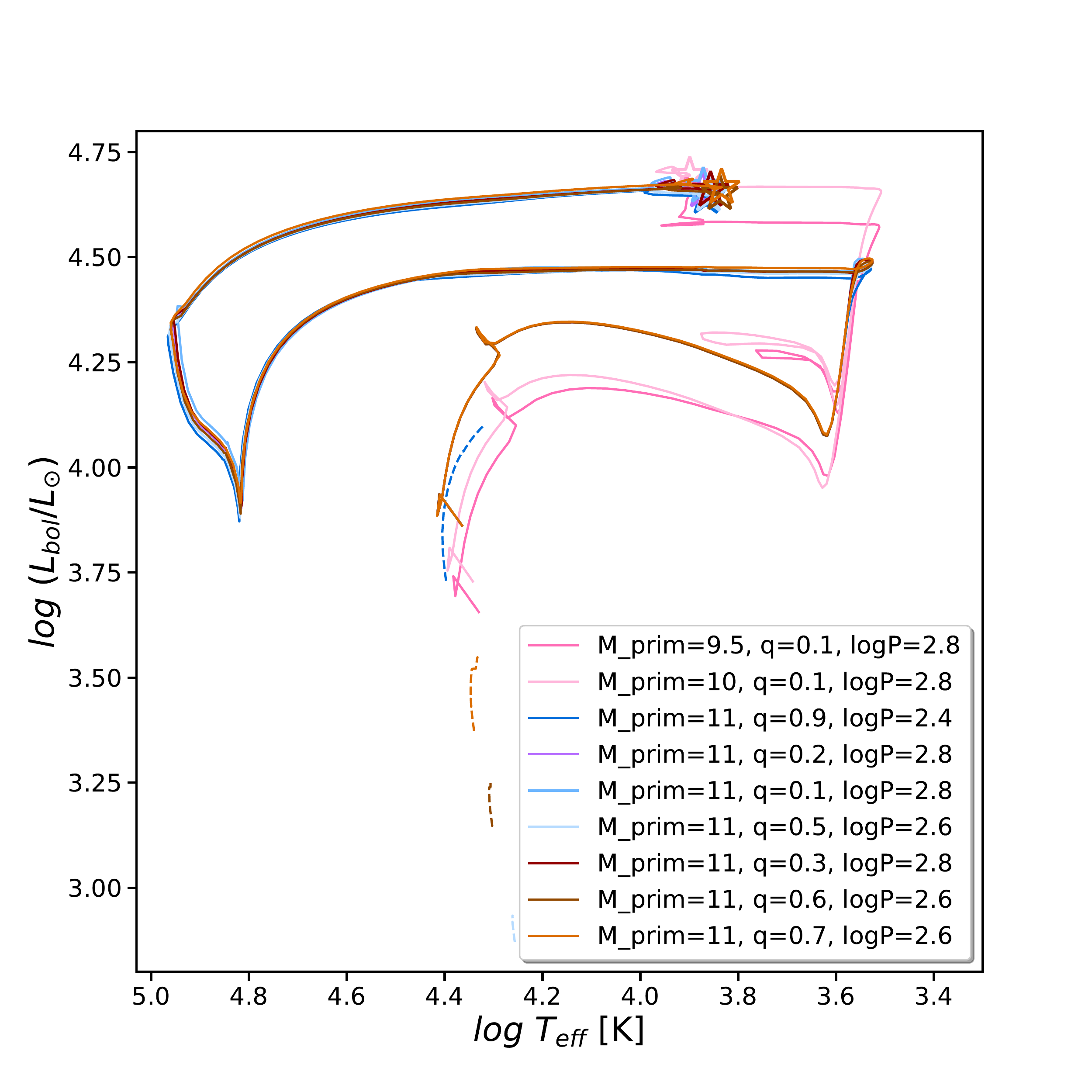}
    \caption{{\it Left}: A colour-magnitude diagram showing the SN 2016gkg progenitor (green filled square), based on our new estimate of its properties (see text). For comparison, we show several model binary systems at the terminus of the primary star from BPASS v2.2.2 (\citealt{Stanway2018}; purple open circles), focusing in particular on those nine models which are consistent with the progenitor locus, to within the uncertainties. See figure legend --- the models are distinguished by the initial mass of the primary, $M_{\rm prim}$; the mass ratio, $q$; and the initial orbital period (in d), $\log P$. {\it Right}: A Hertzsprung-Russell diagram showing the evolutionary tracks for the primary of the nine BPASS models (solid lines), as well as the tracks for several of the corresponding secondaries (companions; dashed lines) --- less luminous companion tracks are not shown. The open stars denote the locus of the terminus for each model primary.}
    \label{fig:sn2016gkg_bpass}
\end{figure*}

If we assume a host-galaxy distance of 18.2\,Mpc (based on the Numerical Action Methods model; \citealt{Shaya2017,Kourkchi2020}) --- note that this is significantly different from the distance assumed by \citet{Bersten2018} and less than that assumed by \citet{Kilpatrick2022} --- and only Galactic foreground extinction, $A_V=0.053$\,mag (as did \citealt{Bersten2018}) with $R_V=3.1$, then the absolute brightnesses of the progenitor would be $M_{\rm F450W} = -6.79 \pm 0.22$ and $M_{\rm F606W} = -6.97 \pm 0.22$\,mag. In Figure~\ref{fig:sn2016gkg_bpass}, we show this locus for the progenitor in a colour-colour diagram, along with (for comparison) the endpoints for BPASS v2.2.2 \citep{Stanway2018} model binary systems which would lead to SNe~IIb, following the prescription of \citet[][ also J.~Eldridge, private communication]{Eldridge2017}. As can be seen from the figure, a number of models compare well, to within the uncertainties in the progenitor brightness (although, in general, these models are all systematically somewhat more luminous), with $M_{\rm ZAMS} \leq 11$ M$_{\odot}$ and a range of component mass ratios $q$; all of the model systems are of long initial orbital periods $\log P$ (i.e., $P \approx 251$--631\,d). The primary in all of these model systems terminates with $T_{\rm eff} \approx 6300$--7900\,K, $\log (L/{\rm L}_{\odot}) \approx 4.65$, $R_{\rm primary} \approx 118$--154\,$R_{\odot}$, and $M_{\rm ejecta} \approx 1.20$--1.45 M$_{\odot}$. The $T_{\rm eff}$, $\log (L/{\rm L}_{\odot})$, and $R_{\rm primary}$ are somewhat cooler, more luminous, and slightly larger (respectively) than those found by \citet{Kilpatrick2022}; however, both that study and this one determined that the primary was less luminous than previous estimates. The range in radii that we determine now is significantly less than those by \citet{Piro2017} and \citet{Bersten2018}, but it is still roughly consistent with that of \citet{Arcavi2017}. The model primaries we find also terminate with $M_H \approx 0.001$--0.004 M$_{\odot}$, consistent with the results from the modeling of SNe~Ib and SNe~IIb by \citet[][ although see \citealt{Gilkis2022}]{Dessart2011}. All of these parameters should be considered upper-limit constraints, since there is bound to have been fainter stellar emission, other than that of Star A, which was blended with the progenitor in the detected candidate; however, we have no means here of determining that beyond much deeper future observations.

We mention here that the most luminous companion (for model [11 M$_{\odot}$, $q=0.9$, $\log P=2.4$]) has $T_{\rm eff} \approx 20,750$\,K and $\log (L/{\rm L}_{\odot})=4.10$ at the terminus of the primary. More importantly for attempts at detecting the companion (e.g., \citealt{Fox2022}), it has absolute magnitudes $-5.57$ and $-5.05$ at F275W and F336W, respectively. At the distance and reddening to SN 2016gkg, these are apparent magnitudes 25.84 and 26.34. In some future companion search, one could reach these brightnesses at S/N = 4 in total exposure times of $\sim 8700$ and $\sim 6200$\,s, respectively, in these two bands\footnote{According to the {\sl HST\/} WFC3/UVIS Exposure Time Calculator, https://etc.stsci.edu/etc/input/wfc3uvis/imaging/.}. Interestingly, \citet{Kilpatrick2022} already set an upper limit of 26.0 AB mag (24.5 Vega mag) at F275W, based on an available 1500\,s Snapshot observation containing SN 2016gkg on 2021 May 14 (GO-16287; PI J.~Lyman).

However, we mention two final notes related to this discussion. First, for SNe~IIb the default approach has been to invoke binary models for their progenitors, whereas the fact that Cassiopeia~A was an SN~IIb \citep[e.g.,][]{Krause2008} and also not a binary at death \citep{Kochanek2018} seem to contradict the need to search for a surviving companion for this SN type. Second, if core-collapse SNe produce dust, as is generally believed and observed for SN~1987A \citep[e.g.,][]{Matsuura2015}, Cas~A \citep[e.g.,][]{DeLooze2017}, and the Crab Nebula \citep[e.g.,][]{Owen2015}, then a binary companion will be obscured by the dust at these early phases (unless one invokes the scenario that \citealt{Kochanek2017} proposed, of a sufficiently hot companion which can suppress dust formation) and also should not be detectable; see \citet{Kochanek2017} for a general discussion of this problem.

\subsection{SN 2017eaw}

SN 2017eaw in NGC 6946 generated significant interest in the community, since it is the tenth historical SN to occur in what has been nicknamed the ``Fireworks Galaxy.'' Extensive optical and infrared follow-up observations of this luminous SN~II-P ensued \citep[e.g.,][]{Tsvetkov2018,Tinyanont2019,VanDyk2019,Szalai2019}. \citet{Kilpatrick2018}, \citet{VanDyk2019}, and \citet{Rui2019} all characterised the progenitor candidate that was identified in several {\sl HST\/} bands, as well as in the shortest-wavelength {\sl Spitzer\/} data. Specifically, \citet{VanDyk2019} measured the progenitor candidate to have $26.40 \pm 0.05$ and $22.87 \pm 0.05$\,mag in F606W and F814W, respectively, on 2016 October 26. Those authors conducted detailed dust radiative-transfer modeling of the candidate's spectral energy distribution (SED) and concluded that the star was a dusty, luminous RSG with $M_{\rm ZAMS} \approx 15$ M$_{\odot}$.

\begin{figure*}
	\includegraphics[width=\columnwidth]{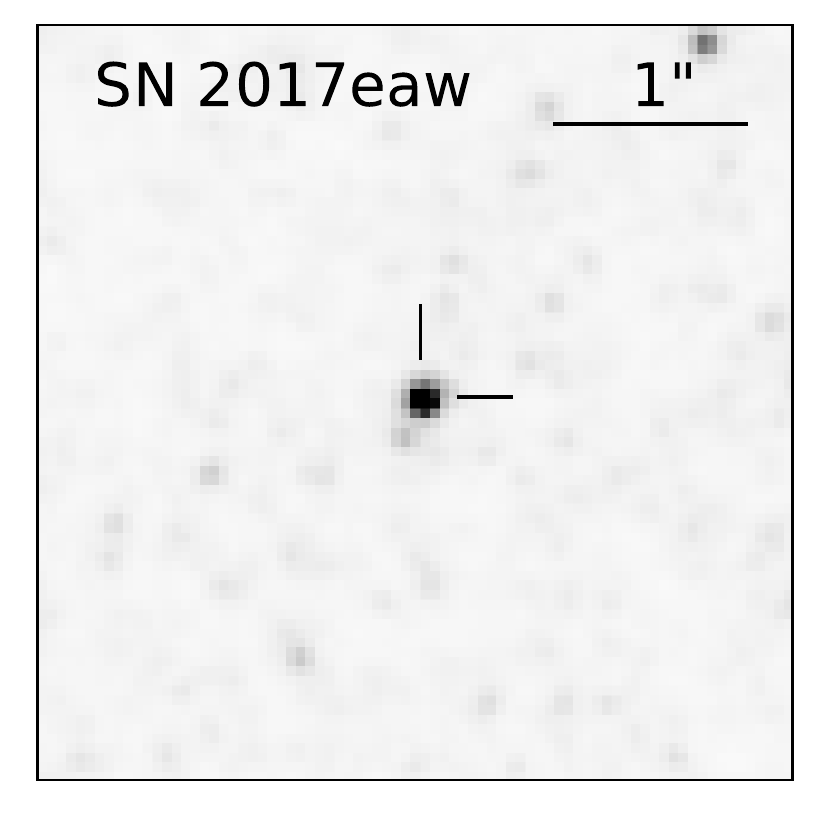}
	\includegraphics[width=\columnwidth]{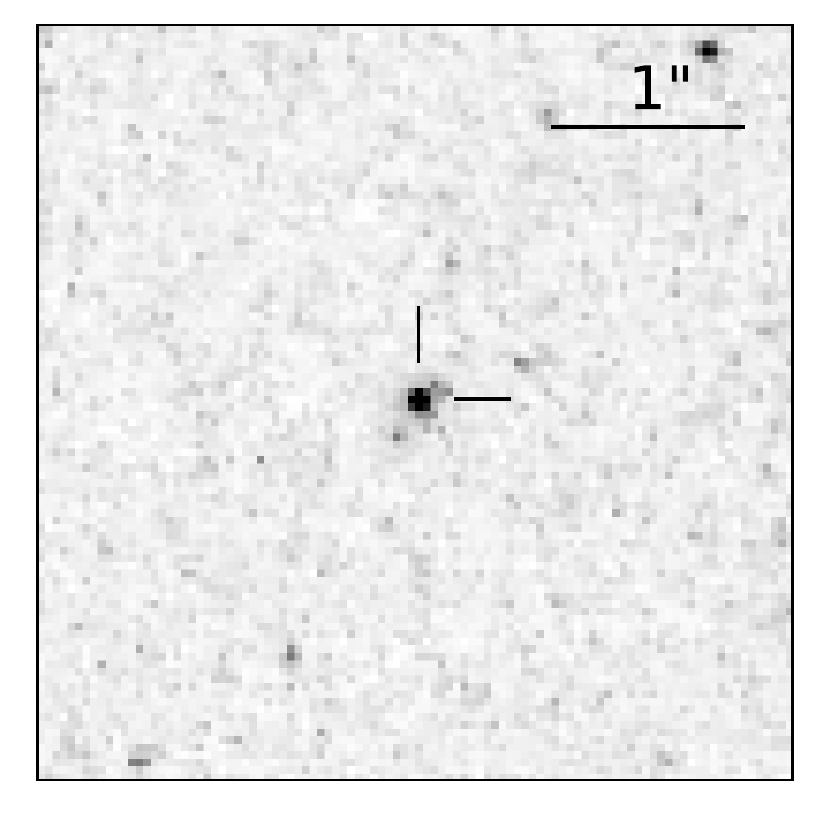}
    \caption{{\it Left}: A portion of the pre-explosion ACS/WFC F814W mosaic from 2016 October 26, with the progenitor candidate for SN 2017eaw indicated by tick marks (see also, e.g., \citealt{VanDyk2019}, their Figure 20). {\it Right}: A portion of the WFC3/UVIS F814W mosaic from 2020 November 11, with the corresponding position of the progenitor candidate indicated by tick marks. North is up, and east is to the left.}
    \label{fig:sn2017eaw}
\end{figure*}

\begin{figure*}
	\includegraphics[width=\columnwidth]{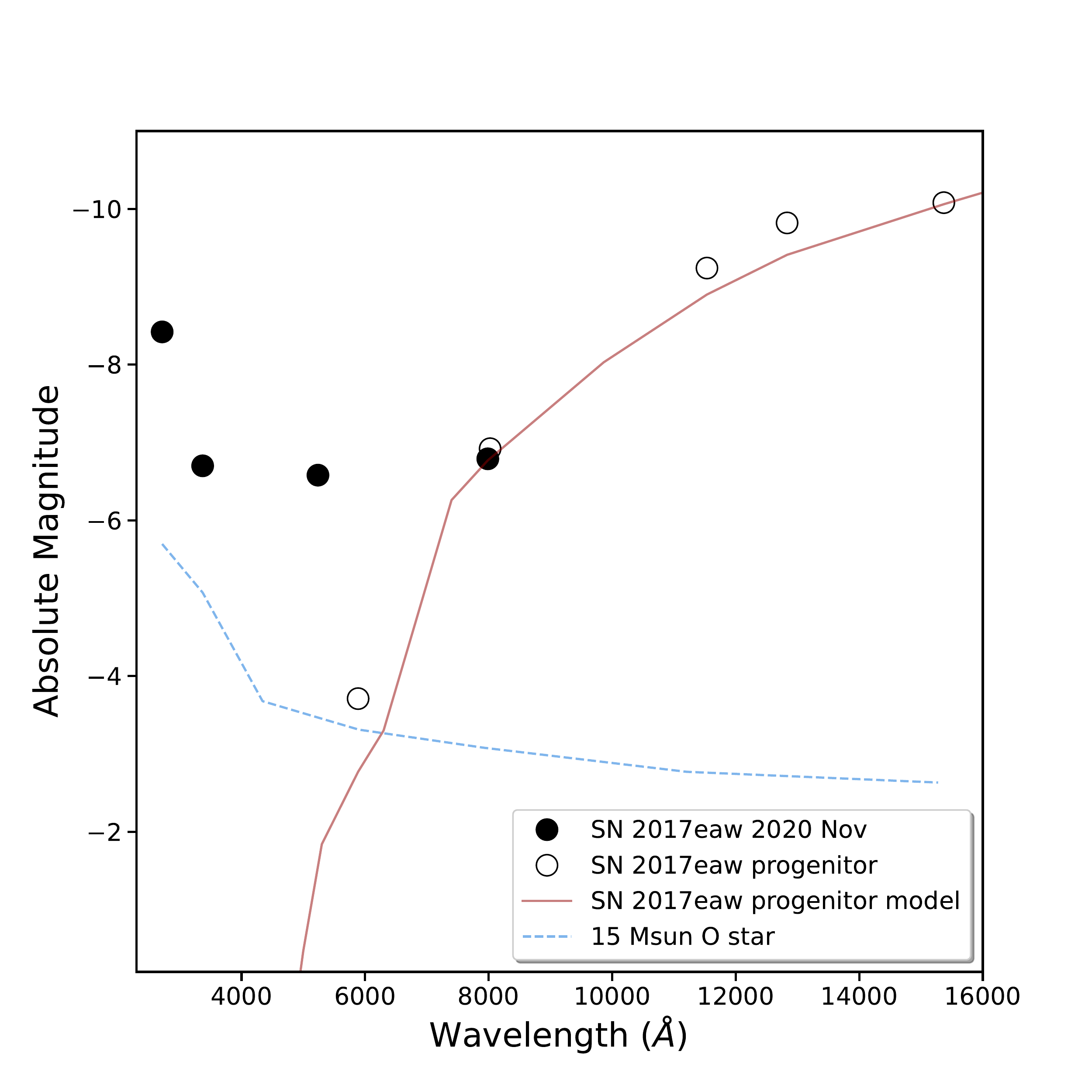}
	\includegraphics[width=\columnwidth]{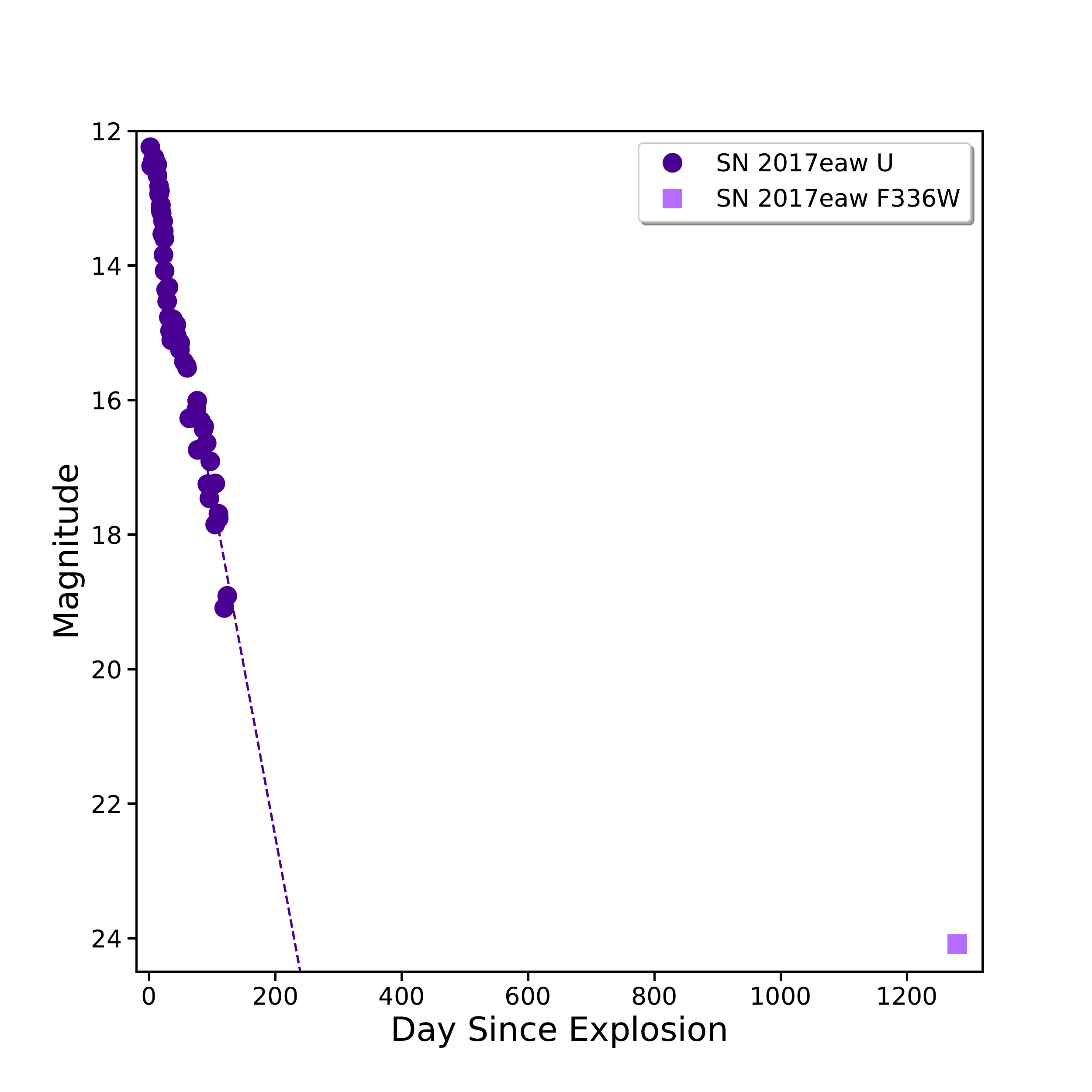}
    \caption{{\it Left}: The SED for SN 2017eaw based on WFC3/UVIS observations from 2020 November (solid circles). Also shown is the SED for the progenitor candidate (open circles), as well as a best-fitting model for the SED (solid curve, adjusted to the distance assumed here for the host galaxy), from \citet{VanDyk2019}. Additionally, we show the SED of a $M_{\rm ZAMS} = 15$ M$_{\odot}$ O-type star at solar metallicity (dashed curve). {\it Right}: The $U$-band light curve of SN 2017eaw (solid circles; \citealt{Tsvetkov2018,Szalai2019}), with an extrapolation in time from the observed light-curve decline (dashed line). Also shown is the F336W data point corresponding to WFC3/UVIS observations at F336W from 2020 November 3 (GO-15877; PI E.~Levesque).}
    \label{fig:sn17eaw_excess}
\end{figure*}

Our Snapshot observations in F555W (710\,s) and F814W (780\,s) are from 2020 November 11, 1279.6\,d (3.5\,yr) after explosion. We show in Figure~\ref{fig:sn2017eaw} the F814W observation compared to the pre-explosion F814W image from 2016 October 26. From these observations we measure $23.84 \pm 0.02$ and $23.17 \pm 0.03$\,mag in F555W and F814W, respectively. For further interpretation here, we assume a distance to the host galaxy of $7.11 \pm 0.38$\,Mpc ($\mu=29.26 \pm 0.12$\,mag), adopting the tip-of-the-red-giant-branch (TRGB) distance estimate from the Extragalactic Distance Database\footnote{http://edd.ifa.hawaii.edu/} (note that this differs from the TRGB distance that \citealt{VanDyk2019} determined and assumed). Following \citet{VanDyk2019}, we assume that the extinction to SN 2017eaw is only due to the Galactic foreground, $A_V=0.941$\,mag (\citealt{Schlafly2011}; with $R_V=3.1$). Referring to the candidate measurements by \citet{VanDyk2019}, the SN at F555W is still $\sim 2.7$\,mag brighter than the progenitor was at F606W (the difference in the bandpasses between WFC3 F555W and ACS/WFC F606W has no effect on this brightness disparity). However, the SN has faded just enough in F814W that we can state it is highly likely that the progenitor candidate has vanished. This is demonstrated graphically in Figure~\ref{fig:sn17eaw_excess}. 

We note that, as the SN has faded and at the higher resolution of WFC3/UVIS, there does appear visually to be faint, extended, diffuse emission, particularly in F555W. The star to the southeast of the SN was obvious in the pre-explosion images, as it is now. However, there are indications in these late-time images in both bands that, at least partially, fainter stars in the immediate environment could be contributing to the extended profile and detected diffuse emission. We can likely exclude a luminous compact light echo, since no obvious indication of scattered early-time SN light appears in the \citet{Weil2020} day 900 SN spectrum (although a low-luminosity echo could still be a component of the late-time emission).

Additionally, we analysed nearly-contemporaneous, publicly-available archival WFC3 data, from 2020 November 3 (only 8.3\,d prior to our Snapshots), in F275W and F336W from GO-15877 (PI E.~Levesque) that serendipitously covered the SN site. From these we measure $22.72 \pm 0.02$ and $24.09 \pm 0.04$\,mag in the two respective short-wavelength bands. Astoundingly, we can see in Figure~\ref{fig:sn17eaw_excess} that the emission at the SN position is significantly brighter in both F336W and F275W than in F555W. We cannot completely rule out, based on the pre-SN F606W and F814W progenitor detections, a pre-existing UV-bright source which is within the PSF of the progenitor; at the distance of the host galaxy, the PSF full width at half-maximum intensity (FWHM) of $\sim$2.1 UVIS pixels ($\sim 0{\farcs}08$) is $\sim$2.8\,pc, so it is conceivable that the profile could contain additional sources within it. For instance, a single O-type star with $M{\rm ZAMS} = 15$ M$_{\odot}$ (i.e., a closely neighbouring massive star at the same turnoff mass as the progenitor) could have been concealed in the light of the progenitor, if we take the F606W and F814W pre-SN measurements at face value (the O-star would have negligible contributions in the IR bands). However, more than one such star would have been harder to reconcile without having substantially more effect on the measured progenitor brightness in the blue. Additionally, what we have detected in the UV/blue at late times is enormously more luminous than a putative O-star neighbour or two; furthermore, the shape of the SED of the UV excess would be highly unusual for a normal star.

We believe that a far simpler explanation for the UV excess seen in 2020 is ongoing late-time SN shock/CSM interaction. This is further illustrated via the $U$-band light curve for SN 2017eaw shown in Figure~\ref{fig:sn17eaw_excess}, with the addition of the 2020 F336W detection, which is many magnitudes more luminous than extrapolation of the decline trend for the early-time $U$ emission, again implying an additional source for the $U$ flux in 2020. The possibility of continued interaction is entirely consistent with that implied by the late-time spectroscopic observations of the SN obtained by \citet{Weil2020} -- the elevated flux at F555W could well be produced by a possible continuation of the strong, boxy H$\alpha$ emission from day 900 to day 1279.6, since that prominent emission would be just within the redward wing of the F555W bandpass. The luminous excess in F275W could be the result of strong Mg~{\sc ii} $\lambda\lambda$2796, 2803 emission, similar to that observed at late times from two extraordinary SNe~IIn, SN 2010jl \citep{Fransson2014} and SN 2005ip \citep{Fox2020}. The slight upturn at F814W could be further enhanced as the result of SN dust, since we know that dust has been forming \citep{Tinyanont2019,Szalai2019}.

As a final note here, we can more solidly confirm that the progenitor has likely vanished, from WFC3 data obtained on 2022 February 12 (GO-16691; PI R.~Foley) with the SN at $23.93 \pm 0.04$ and $23.36 \pm 0.05$\,mag in F555W and F814W, respectively.

\subsection{SN 2018zd}

SN 2018zd in NGC 2146 has been characterised as either a low-luminosity SN~II-P with CSM interaction, from a $M_{\rm ZAMS} \approx 12$ M$_{\odot}$ star which produced a relatively small amount of $^{56}$Ni in the core-collapse explosion \citep{Zhang2020,Callis2021}, or an electron-capture (EC) event from a super-asymptotic giant branch (SAGB) progenitor \citep{Hiramatsu2021}. Much of the source of the debate of SN 2018zd's nature stems from a lack of definitive knowledge of the host galaxy's distance \citep{Callis2021}. The EC scenario has underpinnings, among other aspects, in the identification of a progenitor candidate in {\sl HST\/} image data. \citet{Hiramatsu2021} determined that an object seen pre-explosion at the SN position was likely stellar and had $25.05 \pm 0.15$\,mag in F814W (upper limits to detection were also estimated in F225W and F658N); given the distance they assumed ($9.6 \pm 1.0$\,Mpc), the luminosity was more consistent with an SAGB star than with a more massive RSG. However, should the host be at a much larger distance (e.g., $\sim 18$\,Mpc), the identified progenitor might be more in agreement with an RSG, albeit still at the lower-luminosity end of known SN~II-P progenitors.

\begin{figure*}
	\includegraphics[width=\columnwidth]{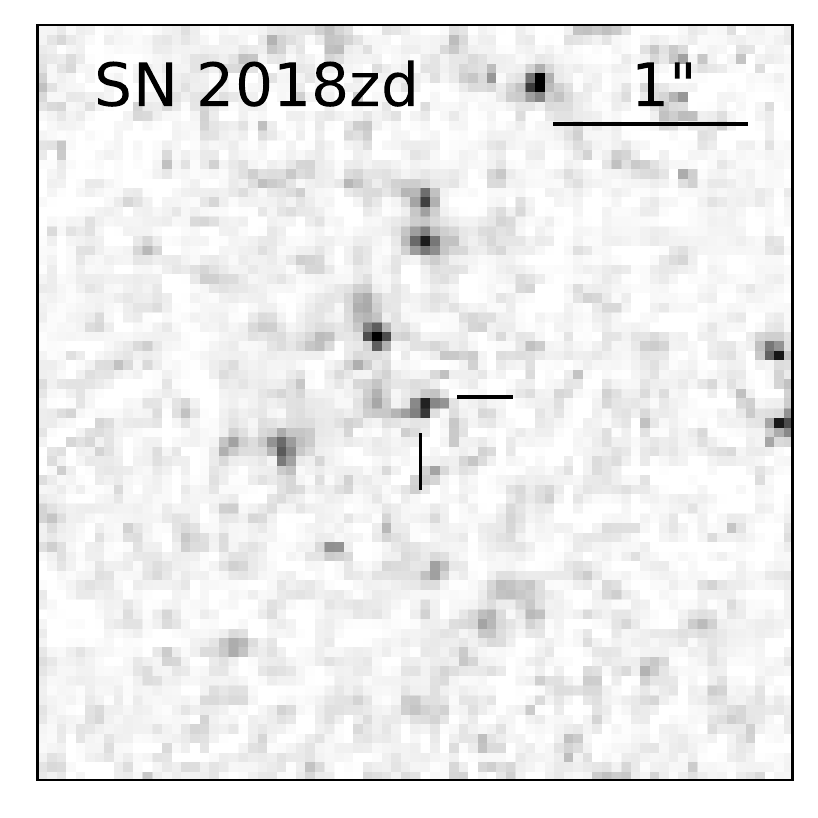}
	\includegraphics[width=\columnwidth]{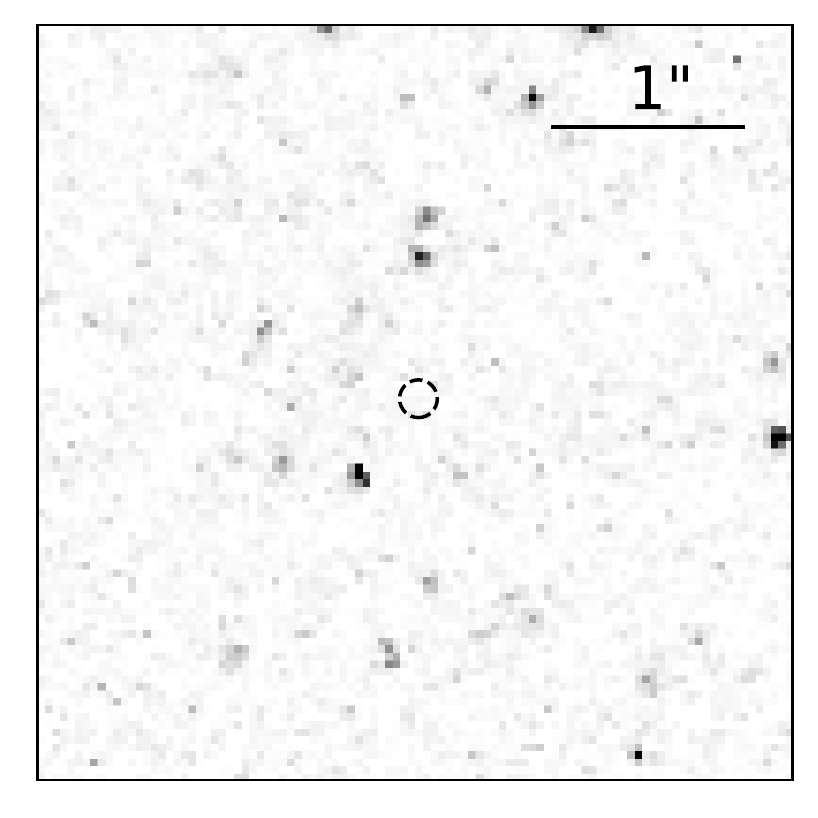}
    \caption{{\it Left}: A portion of the pre-explosion ACS/WFC F814W mosaic from 2004 April 10, with the progenitor candidate for SN 2018zd indicated by tick marks (see also \citealt{Hiramatsu2021}, their Extended Data Figure 1). {\it Right}: A portion of the WFC3/UVIS F814W mosaic from 2021 February 7, with the corresponding position of the progenitor candidate encircled. No source is detected at the SN position to $>26.1$\,mag in that band. Note that the pre-SN image was a single exposure, and CR hits were removed via a deep-learning algorithm \citep{Hiramatsu2021}; the Snapshot mosaic was created from two dithered short exposures. Therefore, some residual CR hits may still be visible in both panels. North is up, and east is to the left.}
    \label{fig:sn2018zd}
\end{figure*}

Our Snapshots were obtained in F606W (710\,s) and F814W (780\,s) on 2021 February 7, 1074.7\,d (2.9\,yr) after explosion. We show the SN site in Figure~\ref{fig:sn2018zd}. Neither the SN nor the precursor object are detected in either band, to $>27.0$ and $>26.1$\,mag, respectively. We therefore consider it likely that the object has vanished and that the candidate, whether an SAGB or RSG, was indeed the progenitor of SN 2018zd. Again, $\gtrsim 1$\,mag of extinction at F814W from freshly-formed dust would be required to obscure the object at the SN position, if it were not directly associated with SN 2018zd.

\subsection{SN 2018aoq}

SN 2018aoq in NGC 4151 is a SN~II-P, possibly less luminous than more-normal events such as SN 1999em, an assessment by \citet{ONeill2019} also supported by overall lower observed ejecta expansion velocities. \citet{Tsvetkov2019} provided further photometric and spectroscopic follow-up data, as well as distance estimates based on the SN itself. \citet{ONeill2019} took advantage of the rich array of multiband, multi-epoch pre-explosion WFC3 data, which were used to measure a Cepheid-based distance to this well-studied Seyfert 1 galaxy \citep{Yuan2020}, to determine the nature of the progenitor candidate isolated in those data. \citet{ONeill2019} had measured an average brightness for the candidate of $26.68 \pm 0.11$ and $23.99 \pm 0.04$\,mag in F555W and F814W, respectively; along with measurements in F350LP and F160W, they concluded that the star was an RSG with $\log (L/{\rm L}_{\odot}) \approx 4.7$ and $M_{\rm ZAMS} \approx 10$ M$_{\odot}$.

\begin{figure*}
	\includegraphics[width=\columnwidth]{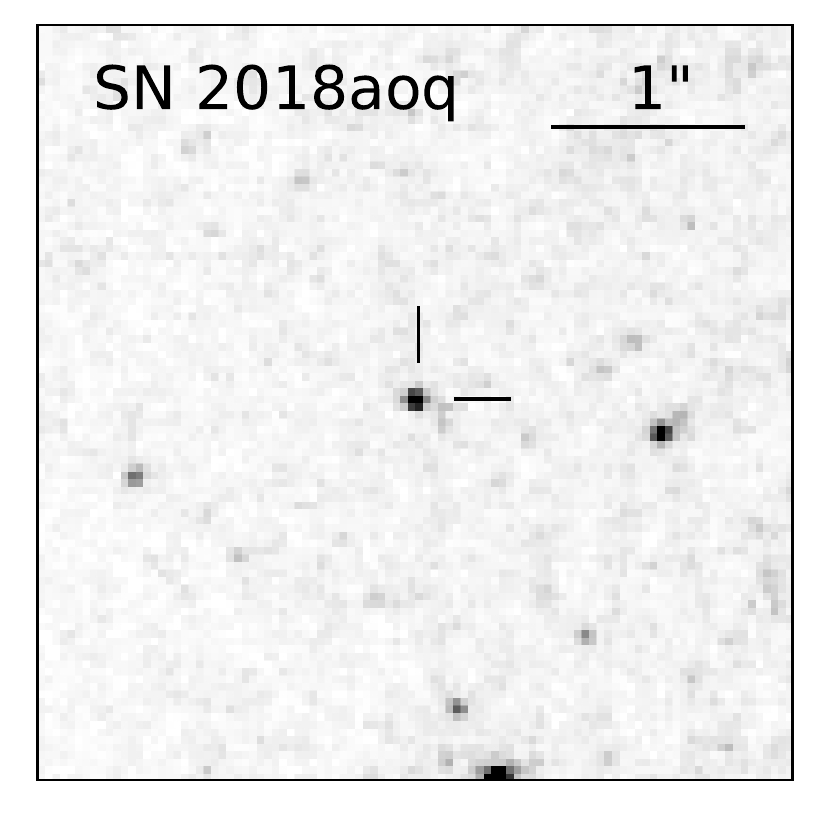}
	\includegraphics[width=\columnwidth]{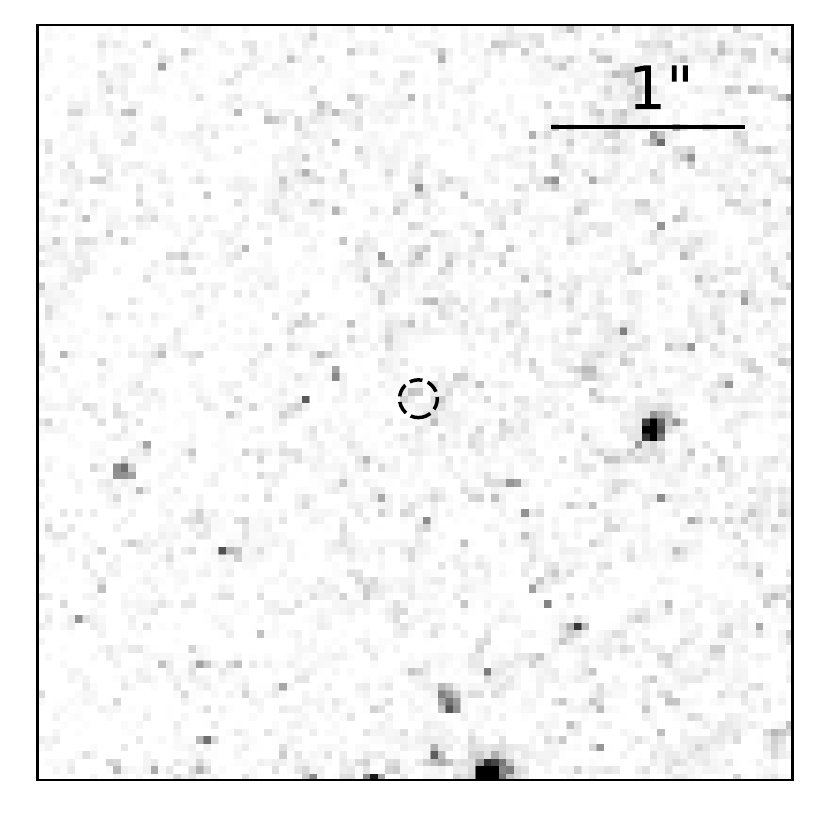}
    \caption{{\it Left}: A portion of the pre-explosion WFC3/UVIS F814W mosaic from 2015 December 21, with the progenitor candidate for SN 2018aoq indicated by tick marks (see also \citealt{ONeill2019}, their Figure 2). {\it Right}: A portion of the WFC3/UVIS F814W mosaic from 2020 December 5, with the corresponding position of the progenitor candidate encircled. No source is detected at the SN position to $>25.9$\,mag in that band. North is up, and east is to the left.}
    \label{fig:sn2018aoq}
\end{figure*}

The Snapshots were obtained in F555W (710\,s) and F814W (780\,s) on 2020 December 5, 981.3\,d (2.7\,yr) post-explosion. We show the image in F814W, compared to the pre-SN image in the same band, in Figure~\ref{fig:sn2018aoq}. We do not detect anything at the SN site to $>26.9$ and $>25.9$\,mag in F555W and F814W, respectively. If the candidate were totally unrelated to the SN progenitor, extinctions of $A_{\rm F555W}$ and $A_{\rm F814W} \gtrsim 1.3$\,mag as a result of dust would be required to make the candidate no longer detectable. It is far more likely, in our opinion, that since lower-luminosity SNe~II-P do not tend to be associated with much post-explosion dust (e.g., \citealt{Meikle2007}), the identified candidate was the progenitor of SN 2018aoq.

\section{Summary}

We have presented the results for six of the 37 visits of nearby SNe from a successfully completed {\sl HST\/} Snapshot program and shown that the progenitor candidates for these CCSNe have likely vanished, confirming these objects as the actual progenitors. This only became possible since each of these SNe (SN 2012A, SN 2013ej, SN 2016gkg, SN 2017eaw, SN 2018zd, and SN 2018aoq) had faded sufficiently in at least one band that the SN has become less luminous than the progenitor candidate. (For SN 2012A, the progenitor candidate was identified in only $K'$ from the ground, not in {\sl HST\/} data as was the case for the other five.) We therefore have added to the list of 17 CCSN progenitors that have been previously confirmed; this increases the current sample size of confirmed CCSN progenitors by about 35\%. It is essential that other SNe whose progenitor has not yet been confirmed be observed with {\sl HST}, or even {\sl JWST}, at sufficiently late times. 

We offered the caveat that CSM dust could be obscuring the progenitor candidate enough that it appears to have disappeared, but may merely have been dimmed by the dust and could still be present. For all but one of the six SNe presented here, we were able to argue that this is most likely not the case. However, we urge further monitoring, with both {\sl HST\/} and {\sl JWST}, of SN 2013ej, for which evidence points to the presence of dust at late times, to confirm our tentative result presented here.

With the addition of {\sl HST\/} F275W and F336W archival data nearly contemporaneous with our Snapshots, we also showed that SN 2017eaw exhibited a UV/blue excess that can best be explained by the existence of ongoing, late-time interaction of the SN with the progenitor's CSM.

With the notable exception of SN 2016gkg, the other five SNe occurred relatively isolated from other stars in their immediate environment, although indications possibly exist of fainter objects in the close vicinity of SN 2017eaw, and we cannot yet successfully distinguish SN 2013ej from a likely luminous blue object near it. Further observations of SN 2017eaw should reveal whether it was a member of a stellar cluster, although the aforementioned CSM interaction could delay such observations from being fruitful for some unknown duration of time. Similar circumstances, as well as the dust mentioned above, could complicate future observations of SN 2013ej.

We also demonstrated that the progenitor candidate originally identified in WFPC2 images of SN 2016gkg likely consisted of a blend of the actual progenitor with a closely neighbouring star, this latter object now becoming more evident in our WFC3 Snapshot data. We then subtracted the light of the neighbour from that of the candidate and find that the progenitor, if the primary in a binary system, likely had at explosion $T_{\rm eff} \approx 6300$--7900\,K, $\log (L/{\rm L}_{\odot}) \approx 4.65$, $R \approx 118$--154\,$R_{\odot}$, and $M_{\rm ZAMS} \approx 9.5$--11 M$_{\odot}$. These parameters represent a revision of estimates made at earlier times in the SN's evolution.

In future attempts at progenitor identification, we therefore recommend that, particularly for SESNe such as SN 2016gkg, any characterization of progenitor properties based on a candidate identified --- especially in pre-explosion WFPC2 images --- should be considered {\em provisional}, until very late-time follow-up observations with WFC3/UVIS are possible. Another salient example of this, although not an SESN, is SN~II 2008cn, for which \citet{Maund2015} (via late-time ACS/WFC imaging) determined that the progenitor candidate, also identified in WFPC2 data \citep{EliasRosa2009,EliasRosa2010}, was actually a blend of two sources, the RSG progenitor and another neighbouring star. Other cases include SN 1999ev \citep{Maund2014}, and SN 2009kr and SN 2009md \citep{Maund2015}, for which each progenitor candidate persisted and had not vanished. In general, this cautionary message is particularly pertinent for SNe occurring in host galaxies at larger distances (e.g., at $\gtrsim 5$--10\,Mpc) and those clearly occurring in relatively crowded environments. 

\section*{Acknowledgements}

We thank the referee for their review, and for providing comments regarding SN~IIb progenitors and detection of their putative companions. We are also grateful to Jay Anderson for a discussion regarding source detection and WFC3 charge-transfer efficiency. This research is based on observations, associated with programs GO-16179 and GO-15877, made with the NASA/ESA {\sl Hubble Space Telescope\/} and obtained from the Space Telescope Science Institute (STScI), which is operated by the Association of Universities for Research in Astronomy, Inc., under NASA contract NAS5-26555. Support for GO-16179 was provided by NASA through a grant from STScI. A.V.F.'s SN team at U.C. Berkeley also received generous support from the Miller Institute for Basic Research in Science (where A.V.F. was a Miller Senior Fellow), Gary and Cynthia Bengier (T.~deJ. was a Bengier Postdoctoral Fellow), the Christopher R.~Redlich Fund, and many individual donors. 

\section*{Data Availability}

All of the {\sl HST\/} data analysed herein are publicly available via the Mikulski Archive for Space Telescopes (MAST) portal, https://mast.stsci.edu/search/ui/\#/hst. All of the photometric results that we have obtained from these data have been listed above. All other data (e.g., BPASS models) are available via other known public sources (e.g., https://bpass.auckland.ac.nz/).


\bibliographystyle{mnras}
\bibliography{ms} 


\bsp	
\label{lastpage}
\end{document}